\documentclass[12pt,preprint]{aastex}

\shorttitle{The Configuration of Neptune Arcs}
\shortauthors{Tsui}
\input{epsf}
\begin{document}
\title{\sc{The Configuration of Fraternite-Egalite2-Egalite1
\\in the Neptune Ring Arcs System}}
\author{K.H. Tsui}
\affil{Instituto de Fisica - Universidade Federal Fluminense
\\Campus da Praia Vermelha, Av. General Milton Tavares de Souza s/n
\\Gragoata, 24.210-340, Niteroi, Rio de Janeiro, Brasil.}
\email{tsui$@$if.uff.br}
\date{}
\pagestyle{myheadings} 
\baselineskip 24pt
\vspace{2.0cm}

\begin{abstract}

By considering the finite mass of Fraternite, although small,
 it is shown that there are two time averaged stationary points
 in its neighborhood due to the reaction of the test body to
 the fields of Neptune, Galatea, and Fraternite. These two
 locations measuring 11.7 and 13.8 degrees from the center of
 Fraternite could correspond to the locations of Egalite 2 and
 Egalite 1. This model accounts for the 10 degree span of
 Fraternite and estimates its mass at $m_{f}=6.4\times 10^{16}$ Kg.
 The eccentricities of Egalite 2 and Egalite 1 are believed
 to be about $e=5\times 10^{-4}$.

\end{abstract}

\vspace{2.0cm}
\maketitle
\newpage

\section{Introduction}

Ever since the discovery of the Neptune arcs [Hubbard et al 1986],
 the constant monitoring of their evolution have revealed much of
 their dynamic properties [Smith et al 1989, Sicardy et al 1999,
 Dumas et al 1999]. Nevertheless, a complete model to account for
 them is still not available. Currently, there are two models that
 attempt to explain their structures. The first is the two-satellite
 model consisting of Galatea for radial confinement of the arcs and
 an hypothetical Lagrange moon for their azimuthal confinement
 [Lissauer 1985, Sicardy and Lissauer 1992]. However, due to the
 Voyager data, a fair size Lagrange moon seems unlikely. The second
 is the one-satellite model where Galatea and the arcs are in the
 42/43 corotation-Lindblad orbit-orbit resonance due to the finite
 eccentricities of Galatea and arcs respectively
 [Goldreich et al 1986, Porco 1991, Foryta and Sicardy 1996].
 However, recent measurements have indicated that the arcs are off
 the corotation resonance location by 0.3 Km and the corotation
 velocity silightly differs from the arc velocity
 [Horanyi and Porco 1993, Sicardy et al 1999, Dumas et al 1999,
 Showalter 1999, Namouni and Porco 2002]. In order to close the
 mean motion mismatch, it is proposed to take into consideration
 the finite mass of the arcs that pulls on the epicycle frequency
 of Galatea [Namouni and Porco 2002]. This would constraint the
 eccentricity of Galatea to residue values which would almost
 eliminate the eccentricity corotation potential although the
 inclination type remains.
 
Further complications come from the angular spreading of and the
 spacing between the arcs. The two minor arcs Egalite 2 and
 Egalite 1 trail behind the main arc Fraternite by about 10 and
 13 degrees respectively measuring from center to center. Fraternite
 has a span of about 10 degrees, Egalite 2 spans over 3 degrees
 while Egalite 1 spans about 1 degree only. Although Fraternite's
 spreading appears to match a 42/43 corotation-Lindblad site of
 eccentricity type, the minor arcs and their spacing from the main
 arc call for inclination resonance or an eccentricity-inclination
 combination [Namouni and Porco 2002]. Here, we attempt to address
 this question of arc spacings. We take the standard eccentricity
 corotation resonance site model to set Fraternite location and
 spreading as a reference position. As for the minor arcs, we
 examine their equations of motion for stationary points over long
 time average. Our model consists of the central body S (Neptune),
 the primary body X (Galatea), a minor body F (Fraternite), and a
 test body s (Egalite 2 and Egalite 1). The center of mass is given
 by SX pair. We consider F and X in corotation-Lindblad resonance
 with each other, and s is coorbital with F. We also take a finite
 mass for F although it is much smaller than that of X. We consider
 gravitational interactions of s and F. Due to the small mass of F,
 only close distance interactions are needed to be considered. By
 expanding in powers of eccentricity of s, and by taking a long
 time average, it is shown that the equations of motion have two
 time averaged stationary points near F that could correspond to
 the positions of the minor arcs.

\newpage
\section{Corotation-Lindblad Resonance}
                     
We designate $M,m_{x},m_{f}$ as the masses of the central body S,
 the primary body X, and the minor body F respectively. Also
 $\vec{r}_{x}=(r_{x},\theta_{x})$, $\vec{r}_{f}=(r_{f},\theta_{f})$,
 and $\vec{r}_{s}=(r_{s},\theta_{s})$ are the position vectors of
 X, F, and s measured from S with respect to a fixed reference axis
 in space. Furthermore, $\vec{R}=\vec{r}_{s}-\vec{r}_{x}$ and
 $\vec{R'}=\vec{r}_{s}-\vec{r}_{f}$ are the position vectors of s
 measured from X and F respectively. We consider all the bodies
 moving on the ecliptic plane by neglecting the orbit inclinations.
 With respect to a coordinate system centered at the central body S,
 the equations of motion of s are
\\
$$\frac{d^2r_{s}}{dt^2}\,
 =\,+\{r_{s}\omega_{s}^2-\frac{GM}{r_{s}^2}\}$$
 
$$-\{\frac{Gm_{x}}{R^3}[r_{s}-r_{x}cos(\Delta\theta_{sx})]
 +\frac{Gm_{x}}{r_{x}^3}r_{x}cos(\Delta\theta_{sx})\}
 -\frac{Gm_{f}}{R'^3}[r_{s}-r_{f}cos(\Delta\theta_{sf})]
 \,\,\,,\eqno(1)$$

$$\frac{1}{r_{s}}\frac{d}{dt}(r_{s}^2\omega_{s})\,
 =\,-\{\frac{Gm_{x}}{R^3}-\frac{Gm_{x}}{r_{x}^3}\}
 r_{x}sin(\Delta\theta_{sx})
 -\frac{Gm_{f}}{R'^3}r_{f}sin(\Delta\theta_{sf})
 \,\,\,.\eqno(2)$$
\\
Here, $\Delta\theta_{sx,sf}=(\theta_{s}-\theta_{x,f})$, whereas
 $\omega_{s}=d\theta_{s}/dt$ is the angular velocity of s about
 the central body S with respect to a reference axis. We expand
 the parameters of s on the right sides of these equations of
 motion in powers of its eccentricity. Taking a time average over
 an interval long compared to the orbital period of s, we have
\\
$$\frac{d^2r_{s}}{dt^2}\,
 =\,(\frac{GM}{L})^2\frac{1}{a}2e^2
 -\frac{m_{x}}{M}(\frac{GM}{L})^2\,a^2\,b_{01}
 +\frac{m_{x}}{M}(\frac{GM}{L})^2\,a^2\,e\,b_{n1}cos(\Phi_{sxL})$$
 
$$+\frac{m_{x}}{M}(\frac{GM}{L})^2\,ar_{x}\,b_{02}
 -\frac{m_{f}}{M}(\frac{GM}{L})^2\,
 \frac{a^2}{R'^3}[1-cos(\Delta\theta_{sf})]\,\,\,,$$

$$\frac{1}{r_{s}}\frac{d}{dt}(r_{s}^2\omega_{s})\,
 =\,-\frac{m_{x}}{M}(\frac{GM}{L})^2\,ar_{x}\,2e\,b_{n2}
 sin(\Phi_{sxL})
 -\frac{m_{f}}{M}(\frac{GM}{L})^2\,
 \frac{a^2}{R'^3}sin(\Delta\theta_{sf})\,\,\,,$$
\\
where $L^2=GMa$ with $a$ as the semi-major axis of s, and
 $\Phi_{sxL}=[(n+1)\theta_{s}-n\theta_{x}-\phi_{s}]$ is the
 Lindblad resonance variable of s with s and F outside of X.
 The coefficients $b_{01}$, $b_{n1}$, $b_{02}$, and $b_{n2}$
 are defined through the Laplace coefficients as
 $b_{01}=(1/2)(1/a^3)b^{(0)}_{3/2}$,
 $b_{n1}=(1/2)(1/a^3)b^{(n)}_{3/2}$,
 $b_{02}=(1/2)(1/a^3)b^{(1)}_{3/2}$,
 $b_{n2}=(1/4)(1/a^3)[b^{(n+1)}_{3/2}+b^{(n-1)}_{3/2}]$.

Rewriting $\Phi_{sxL}=[\Phi_{fxL}+(n+1)\Delta\theta_{sf}
 -(\phi_{s}-\phi_{f})]$ in terms of the FX corotation-Lindblad
 resonance variable $\Phi_{fxL}$, and taking $\phi_{s}=\phi_{f}$,
 the above equations become
\\
$$\frac{d^2r_{s}}{dt^2}\,
 =\,(\frac{GM}{L})^2\frac{1}{a}
 \{2e^2-\frac{m_{x}}{M}a^3\,[b_{01}-\frac{r_{x}}{a}b_{02}]$$
 
$$+\frac{m_{x}}{M}a^3\,e\,b_{n1}\,
 cos[\Phi_{fxL}+(n+1)\Delta\theta_{sf}]
 -\frac{m_{f}}{M}\frac{a^3}{R'^3}[1-cos(\Delta\theta_{sf})]\}
 \,\,\,,\eqno(3)$$

$$\frac{1}{r_{s}}\frac{d}{dt}(r_{s}^2\omega_{s})\,
 =\,-(\frac{GM}{L})^2\frac{1}{a}
 \{\frac{m_{x}}{M}a^2r_{x}\,2e\,b_{n2}\,
 sin[\Phi_{fxL}+(n+1)\Delta\theta_{sf}]
 +\frac{m_{f}}{M}\frac{a^3}{R'^3}\,sin(\Delta\theta_{sf})\}
 \,\,\,.\eqno(4)$$
\\
The angular positions of s where the time averaged force acting
 on it vanishes are given by
\\
$$2e^2+\frac{m_{x}}{M}a^3\,e\,b_{n1}\,
 cos[\Phi_{fxL}+(n+1)\Delta\theta_{sf}]$$
 
$$-\frac{m_{x}}{M}a^3\,[b_{01}-\frac{r_{x}}{a}b_{02}]
 -\frac{m_{f}}{M}\frac{a^3}{R'^3}
 [1-cos(\Delta\theta_{sf})]\,=\,0
 \,\,\,,\eqno(5)$$

$$e\,=\,-\frac{m_{f}}{m_{x}}\frac{a^3}{R'^3}
 \frac{1}{a^2r_{x}\,2b_{n2}}
 \frac{sin(\Delta\theta_{sf})}
 {sin[\Phi_{fxL}+(n+1)\Delta\theta_{sf}]}
 \,\,\,.\eqno(6)$$
\\
These are the general conditions for vanishing time averaged
 force for two coorbital objects s and F that are in corotation
 orbital resonance with an interior X.

\newpage
\section{Fraternite-Egalite2-Egalite1}

Let us now apply these conditions to the Neptune ring arcs.
 With the Neptune system parameters, the last term on the left
 side of the first equation can be neglected unless $R'/a$ is
 exactly zero which amounts to a collision. Also, considering
 the mass ratio $m_{x}/M$ of the Neptune system much less than
 the expected eccentricity such that the linear eccentricity term
 can be neglected. We, therefore, keep only the quadratic term
 and the stationary locations are given by
\\
$$(\frac{R'}{a})^3\,
 \frac{sin[\Phi_{fxL}+(n+1)\Delta\theta_{sf}]}
 {sin(\Delta\theta_{sf})}\,
 =\,-\frac{1}{2}\frac{m_{f}}{m_{x}}
 \frac{a}{r_{x}}\frac{1}{a^3}\frac{1}{b_{n2}}
 \{2\frac{M}{m_{x}}\frac{1}{a^3}
 \frac{1}{[b_{01}-(r_{x}/a)b_{02}]}\}^{1/2}\,\,\,.\eqno(7)$$
\\
Writing $(R'/a)=2\,sin(\Delta\theta_{sf}/2)$ Eq.(7) reads
\\
$$(2\,sin(\frac{\Delta\theta_{sf}}{2}))^3\,
 \frac{sin[\Phi_{fxL}+(n+1)\Delta\theta_{sf}]}
 {sin(\Delta\theta_{sf})}\,
 =\,-\frac{1}{2}\frac{m_{f}}{m_{x}}
 \frac{a}{r_{x}}\frac{1}{a^3}\frac{1}{b_{n2}}
 \{2\frac{M}{m_{x}}\frac{1}{a^3}
 \frac{1}{[b_{01}-(r_{x}/a)b_{02}]}\}^{1/2}\,\,\,.$$
\\
With $\alpha=r_{x}/a=0.98444$, the Laplace coefficients are
 $b^{(0)}_{3/2}=0.26487\times 10^4$,
 $b^{(1)}_{3/2}=0.26470\times 10^4$,
 $b^{(41)}_{3/2}=0.20168\times 10^4$,
 $b^{(42)}_{3/2}=0.19975\times 10^4$,
 $b^{(43)}_{3/2}=0.19782\times 10^4$.
 The right side of the above equation can be calculated to give
\\
$$(2\,sin(\frac{\Delta\theta_{sf}}{2}))^3\,
 \frac{sin[\Phi_{fxL}+(n+1)\Delta\theta_{sf}]}
 {sin(\Delta\theta_{sf})}\,
 =\,-1.5528\times 10^{-4}\frac{m_{f}}{m_{x}}
 (\frac{M}{m_{x}})^{1/2}\,
 =\,-1.1\frac{m_{f}}{m_{x}}\,\,\,.\eqno(8)$$
\\
The second equality is reached by taking $M=1\times 10^{26}$ Kg
 for the central body Neptune, and $m_{x}=2\times 10^{18}$ Kg
 for the primary body Galatea. This leaves only one parameter
 $m_{f}/m_{x}$ in the equation.
 
Considering the center of Fraternite be at the maximum of the
 corotation site with $\Phi_{fxL}=\pi/2$, the positions where
 the time averaged force vanishes are given by 
\\
$$(2\,sin(\frac{\Delta\theta_{sf}}{2}))^3\,
 \frac{cos[(n+1)\Delta\theta_{sf}]}
 {sin(\Delta\theta_{sf})}\,
 =\,-1.1\frac{m_{f}}{m_{x}}\,\,\,.\eqno(9)$$
\\
On the left side of Eq.(9), the cosine function starts with a
 central maximum at $(n+1)\Delta\theta_{sf}=0$ and reaches its
 first minimum at $(n+1)\Delta\theta_{sf}=\pm\pi$ on each side
 forming a complete site of of 8.4 degrees with $n=42$. However,
 due to the other factor, the central maximum is replaced by a
 null and a nearby maximum on each side of it. Numerical solution
 of Eq.(9) in Fig.1 shows the first minimum slightly shifted
 outwards to 4.85 degrees on each side spaning an angular width
 of 9.7 degrees which corresponds to the observed extension of
 Fraternite. The second minimum is located at 12.8 degrees from
 the center. The roots of Eq.(9) are given by the intercepts of
 the left side with the right side. There are either two intersects
 around this second minimum or non of them. Taking the mass ratio
 $m_{f}/m_{x}=3.2\times 10^{-2}$ gives two intercepts at 11.7 and
 13.8 degrees which are approximately where the minor arcs are
 observed. This intercept corresponds to a mass ratio
 $m_{f}/m_{x}=3.2\times 10^{-2}$ which gives
 $m_{f}=6.4\times 10^{16}$ Kg for Fraternite. The slight difference
 of one degree or so between the calculated and observed positions
 is probably because we have represented the elongated distribution
 of Fraternite's mass by a point mass at its center. Besides giving
 a mass estimate of Fraternite and the positions of Egalite 2 and
 Egalite 1, we can also estimate the eccentricity of the two minor
 arcs by using Eq.(5) or Eq.(6). Simple calculations from both
 equations give $e=5\times 10^{-4}$ approximately.

\newpage
\section{Conclusions}

To conclude, we have studied a four-body system where the center
 of mass is set by the central and primary bodies. A minor body
 is in corotation-Lindblad resonance with the primary body, and
 a test body is coorbital with the minor body at close distances.
 Through the equations of motion, we have shown that there are
 stationary locations where the time averaged force vanishes.
 These points are located behind the minor body as well as in front
 of it. They differ from the Lagrangian points of a restricted
 three-body system in that the averaged force is zero, and that
 they are dynamically self-generated by the test body's reaction
 to the fields of the minor and primary bodies in orbit-orbit
 resonance plus the field of the central body. We have applied
 these points to account for the arcs' configuration in the
 Neptune-Galatea system. Using this model, it is able to explain
 the 10 degree extension of Fraternite. By requiring Fraternite's
 mass be $6.4\times 10^{16}$ Kg, two locations with vanishing time
 averaged force exist at 11.7 and 13.8 degrees from the center of
 Fraternite which seem to be compatible with the observed positions
 of Egalite 1 and Egalite 2. It also estimates the eccentricity of
 Egalite 2 and Egalite 1 at $5\times 10^{-4}$.

\centerline{\bf Acknowledgments}

This work was supported by the Conselho Nacional de Desenvolvimentos
 Cientifico e Tecnologico (CNPq, The Brazilian National Council of
 Scientific and Technologic Developments) and the Fundacao de Amparo a
 Pesquisa do Estado do Rio de Janeiro (FAPERJ, The Research Fostering
 Foundation of the State of Rio de Janeiro).

\clearpage
\begin{figure}
\plotone{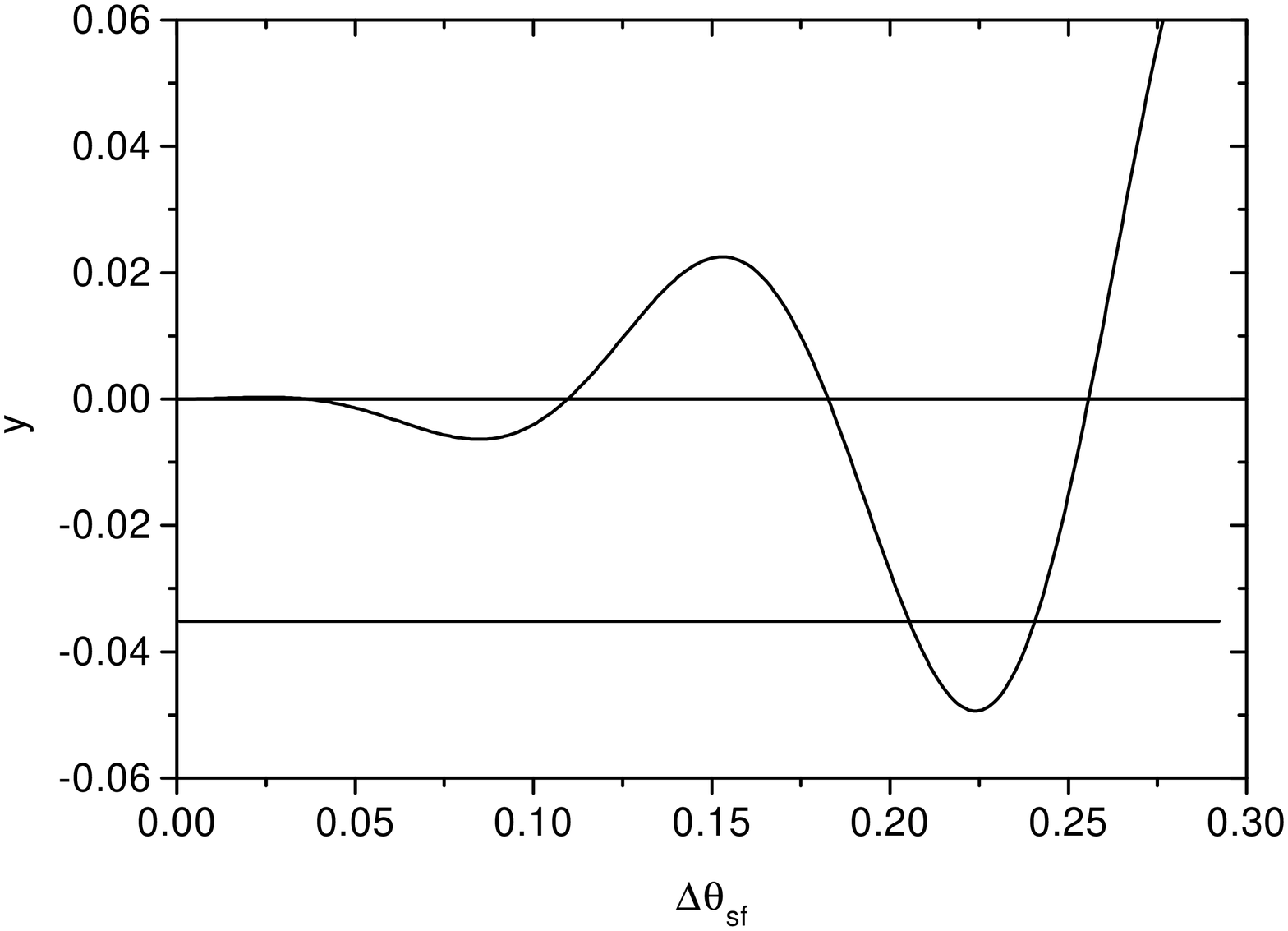}
\caption{Denoted by the y label, the left and right sides of Eq.(9)
 are plotted as a function of $\Delta\theta_{sf}$ in rad/s,
 and the intercepts define the locations where the time averaged
 force vanishes.\label{fig.1}}
\end{figure}

\end{document}